\documentclass[oldversion]{aa} 
\usepackage{graphicx}
\usepackage{longtable}
\usepackage{txfonts}
\usepackage{rotating}
\usepackage{lscape}
\newcommand{\gsim}{\;\lower.6ex\hbox{$\sim$}\kern-7.75pt\raise.65ex\hbox{$>$}\;}
\newcommand{\lsim}{\;\lower.6ex\hbox{$\sim$}\kern-7.75pt\raise.65ex\hbox{$<$}\;}

\begin{document}
\title{The Na-O anticorrelation in horizontal branch stars. I. NGC~2808
\thanks{Based on observations collected at 
ESO telescopes under programme 386.D-0086}
}

\author{
R.G. Gratton\inst{1},
S. Lucatello\inst{1},
E. Carretta\inst{2},
A. Bragaglia\inst{2},
V. D'Orazi\inst{1},
\and
Y. Al Momany\inst{1,3}}

\authorrunning{R.G. Gratton}
\titlerunning{Na-O in HB stars of NGC~2808}

\offprints{R.G. Gratton, raffaele.gratton@oapd.inaf.it}

\institute{
INAF-Osservatorio Astronomico di Padova, Vicolo dell'Osservatorio 5, I-35122
 Padova, Italy
\and
INAF-Osservatorio Astronomico di Bologna, Via Ranzani 1, I-40127, Bologna, Italy
\and
European Southern Observatory, Alonso de Cordova 3107, Vitacura, Santiago, Chile }

\date{}
\abstract{Globular clusters have been recognized to host multiple stellar
populations. A spectacular example of this is the massive cluster NGC~2808,
where multiple populations have been found along the horizontal branch (HB) and
the main sequence (MS). Studies of red giants showed that this cluster
appears homogeneous insofar Fe abundance is concerned, but it shows an
extended anticorrelation between Na and O abundances. The Na-poor, O-rich
population can be identified with the red MS, and the Na-rich,
O-poor one with the blue one. This may be
understood in terms of different He content, He being correlated with Na. 
A prediction of this scenario is that He-rich, Na-rich He-core burning stars, 
because they are less massive, will end up on the bluer part of the HB,
while He-poor, Na-poor stars will reside on the red HB.
The aim of this paper is to verify this prediction. To this purpose, we
acquired high-resolution spectra of regions including strong O and Na lines
in several tens of HB stars of NGC~2808, sampling both
the red and blue parts of the HB. We limited our analysis to those blue HB
stars cooler than the gap at 11,500 K, because diffusion and radiative pressure 
are known to strongly modify the atmospheric composition of warmer stars. 
We indeed found a strict correspondence between the colour of the HB stars and 
their Na and O abundances: all blue HB stars are very O-poor and Na-rich. In
addition, we found that while all the red HB stars are more O-rich and Na-poor 
than the blue ones, there is a moderate Na-O anticorrelation among them as well. 
This anticorrelation is in turn related to the colour of the red HB
stars. These results reinforce the connection between Na and O abundances and
the second parameter phenomenon, and show that there are more than three stellar 
populations in NGC~2808 because only a fraction of the red HB stars belong
to the primordial population of this cluster.}

\maketitle

\section{Introduction}

In the recent years, it has become clear that the formation of the most
massive stellar clusters, the globular clusters (GCs), is a complex phenomenon,
involving several episodes of star formation, with the latest generations
of stars forming from material polluted by the ejecta of a fraction of
the earliest (primordial) population, as first suggested by Norris et al.
(1981; see also Gratton et al. 2001, 2004; Carretta
et al. 2010). Evidence for the presence of these different populations
stems from their chemical composition, in particular the abundances of the
light elements involved in $p-$capture processes (C, N, O, Na, Mg, Al, Si:
Osborn 1971; see also Cottrell \& Da Costa 1981; Kraft 1994;
Denisenkov and Denisenkova 1989; Carretta et al. 2009a, 2009b, and 
references therein), as well as from
the splitting of sequences in the colour magnitude diagram (see e.g. Piotto 2008).
While in the recent years some spectacular splitting of the main sequence (MS) have
been found (see e.g. Piotto et al. 2007), the most dramatic evidence concerns 
the horizontal branch (HB) and was found already
several decades ago (Harris 1974). This splitting of the HB has long been related
to the second parameter problem (Sandage \& Wildey 1967, van den Bergh 1967) but it 
was only quite recently that a plausible cause has been identified in variations 
of the He content related to the multiple populations (D'Antona \& Caloi 2004;
but see Norris et al. 1981 for a very similar early suggestion). 
As pointed out by D'Antona et al. (2002), variations in the He content among 
different stellar generations in GCs are expected to be present, correlated to 
the variation of the abundances of the $p-$capture process elements. These He 
abundance variations are invoked to explain the multiple MSs seen in mono-metallic 
GCs like NGC~2808. However, because He-rich stars are expected to burn H at a 
faster rate than the He-poorer ones while on the MS, their progeny 
on the HB is expected to be less massive, then bluer, than the progeny of 
He-normal stars. Carretta et al. (2007) and Gratton et al. (2010) discussed 
the general correlation between the abundances of $p-$capture elements and the 
colour of HB stars, and found indeed that a close relation exists between the 
extension of the Na-O and Mg-Al anticorrelations, and that of the HB, precisely 
as expected in this interpretative scenario. These studies also showed that the 
main parameter driving the whole phenomenon of multiple populations is the 
cluster mass (confirming previous similar results obtained by Recio-Blanco et 
al. 2006; Carretta et al. 2006, 2009a), and that the second parameter can be 
explained almost entirely by a combination of variations of the age and mass of GCs. 

While this is extremely intriguing, a definitive confirmation of this scenario with
direct determination of the chemical composition of HB stars is required.
The surface composition of the hottest stars on the HB, those with
temperatures $>11,500$~K (the location of the so-called $u-$jump: Grundahl et al.
1999), are known to be heavily influenced by sedimentation caused by diffusion 
and by levitation through radiative pressure effects (see, e.g., Behr et al. 1999, 2003; 
see also Grundahl et al. 1999; Caloi 1999). However, the same studies revealed 
that abundances for cooler stars can be reliably determined, although the low 
$S/N$\ and wavelength range of the spectra available at the epoch prevented an accurate 
determination of the abundances of those elements of interest here (O, Na).
Behr et al. (2003) first obtained He abundances for cool BHB stars but the large 
error bars did not allow the authors to use them as diagnostics for the multiple population
scenario. Very recently, a few investigations tried to measure the abundances of 
O and Na, and even of He, in such stars with sufficiently good accuracy. The pioneering 
study was the analysis of five blue HB stars 
in NGC~6752 by Villanova et al. (2009). This cluster has a very blue HB, and the 
observed stars (with temperatures $>11,500$~K) are among the coolest HB stars. They 
behave as expected, most of them being Na-poor and O-rich, save for one star, which 
is more luminous than the others, and likely is an evolved object that probably 
started its HB evolution at a much higher temperature. Two other studies
(Marino et al. 2011 and Villanova et al. 2011) focused on M~4, the nearest GC. 
This cluster has a much redder HB than NGC~6752, so that stars on both the red 
and blue sides of the RR Lyrae instability strip could be studied. Again, as 
expected, the red HB stars were found to be Na-poor and O-rich, and the blue ones 
Na-rich and O-poor. 

In this paper we present a similar investigation of the HB of NGC~2808.
NGC~2808 is among the most massive Galactic GCs, and has become the most
classical example of multiple populations. For a long time it was
known to have a very peculiar HB, with a multimodal distribution of colours
and masses (Harris 1974; see Bedin et al. 2000,  Momany et al. 2004, and 
Dalessandro et al. 2011 for recent
accurate photometry). D'Antona et al. (2004, 2005) observed a widening of the MS,
later found by Piotto et al. (2007) to consist of three distinct sequences,
which can be explained with different values of the He content. D'Antona et
al. (2004) correlated the different MSs with the different
populations of HB stars. Low S/N, high-resolution spectra along the BHB
of NGC~2808 were obtained by Recio-Blanco et al. (2004) and were used to derive radial and
rotational velocities. Pace et al. (2006) obtained intermediate-resolution 
spectra for a number of HB stars, but they were only able to derive
the Fe abundances from these spectra. They found that stars cooler than
the $u-$jump at $\sim 12,000$~K provide values similar to those obtained
for red giants, while metallicity increases
with temperature for stars warmer than this limit. Carretta et al. (2004, 2006) 
studied the Na-O anticorrelation in several hundred red giants, and found an
extended anticorrelation, with different peaks in the distribution. 
Bragaglia et al. (2010a) used this same observational material to derive
evidence of the variation of the He content along the RGB, with results
that on the whole agree with those obtained from the colours of
MS and HB. However, while very suggestive, the connection between 
the chemical anomalies along the RGB and the multiple MS and HB could not be 
definitively shown by investigations that used very different techniques
on stars in different evolutionary phases. To provide this definitive
evidence, about one year ago, Bragaglia et al. (2010b) obtained spectra for 
one star on the blue MS, and another on the red one. They found 
that as expected, the blue MS star is Na- and Al-rich and Mg-poor, 
while the red one is Na- and Al-poor and Mg-rich. In this paper, we intend to
complete the observational pattern by determining the abundances of O and Na
for several tens of HB stars, using spectra acquired with FLAMES+GIRAFFE at VLT.
We noticed immediately that while we possibly detected the He I line at 
5876~\AA\ in the spectra of several of the BHB stars, stars are so faint that 
only prohibitively long exposure times would have ensured the S/N required 
for accurate measure of the He abundance, as needed to distinguish
primordial abundances from those expected for second generation stars in the
multiple population scenarios\footnote{In practice, among stars with temperature
$<11,500$~K, the He~I line is detected only for star \#32324, and even in that 
case with such an error on the equivalent width that the estimate on He abundance 
is not useful. For the coolest stars the He line is not detectable in our noisy spectra
even when they were summed together.} Therefore we limited our analysis to Na and 
O abundances, for which 
the S/N required to provide sensible results is much lower.

The structure of the paper is the following: in Section 2 we present the 
observational data; in Section 3 we explain our analysis methods; in Section 4
we discuss the Na-O anticorrelation; conclusions are drawn in Section 5.

\begin{table*}[htb]
\centering
\caption[]{Basic data for program stars}
\begin{scriptsize}
\begin{tabular}{cccccccccc}
\hline
  Star & RA (J2000) & Dec (J2000) & $B$ & $V$ & $K$ & $V_r$& S/N& S/N& Note \\
       &            &             &(mag)&(mag)&(mag)&(km/s)&HR12&HR19&      \\
\hline
\multicolumn{10}{c}{Blue HB}\\
\hline
  7136 & 9~11~49.831 &-64~49~35.89 &17.360 &17.193 &15.671 &       & ~7.7 &      & \\
  7189 & 9~12~53.801 &-64~49~32.72 &16.719 &16.544 &16.251 &  106  & 16.0 & 11.6 & \\ 
  8008 & 9~12~49.328 &-64~49~04.85 &17.241 &17.074 &       &  117  & ~9.7 & ~7.5 & \\ 
  9868 & 9~12~01.742 &-64~47~34.54 &16.697 &16.664 &       &       & 16.5 & 10.0 & \\
 10449 & 9~12~17.199 &-64~46~40.87 &16.982 &16.803 &15.433 &       & 13.4 &      & \\ 
 10719 & 9~12~24.790 &-64~46~09.95 &17.906 &17.795 &       &       & ~7.5 & ~2.5 & \\
 14516 & 9~11~37.245 &-64~47~27.86 &16.775 &16.660 &16.073 &       & 15.0 & 13.9 & \\
 14598 & 9~10 23.139 &-64~47~15.20 &17.000 &16.904 &       &       & 12.8 & ~6.5 & \\
 14923 & 9~11~27.828 &-64~46~22.48 &17.582 &17.471 &       &       & ~7.6 & ~4.3 & \\ 
 15924 & 9~11~41.884 &-64~42~05.67 &17.894 &17.975 &       &       & ~6.2 & ~4.3 & \\
 31470 & 9~11~42.487 &-64~54~01.04 &17.337 &17.250 &       &       & ~5.9 & ~7.1 & \\ 
 32324 & 9~11~36.900 &-64~53~05.85 &17.219 &17.076 &       &  112  & 12.3 & ~8.9 & \\ 
 36840 & 9~12~31.524 &-64~58~53.27 &18.149 &18.126 &       &       & ~4.8 & ~1.6 & \\
 37288 & 9~12~34.354 &-64~57~33.47 &17.722 &17.691 &       &       & ~4.2 & ~2.3 & \\
 37289 & 9~11~50.141 &-64~57~33.78 &16.954 &16.797 &15.427 &   95  & ~9.4 & ~7.9 & \\ 
 37345 & 9~12~12.465 &-64~57~26.97 &18.033 &17.916 &       &       & ~3.3 & ~3.6 & \\
 37456 & 9~11~58.235 &-64~57~13.01 &17.049 &16.917 &16.127 &       & 11.5 & ~5.2 & \\ 
 40169 & 9~11~44.754 &-64~54~44.00 &17.233 &17.133 &       &       & 11.8 & ~7.8 & \\ 
 40309 & 9~12~02.205 &-64~54~39.87 &17.291 &17.204 &       &       & 14.0 & ~5.9 & \\
 41077 & 9~12~18.701 &-64~54~17.79 &16.939 &16.824 &15.946 &       & 12.0 & ~6.1 & \\ 
 41388 & 9~12~10.354 &-64~54~10.58 &16.993 &16.884 &16.192 &       & 10.1 & ~9.2 & \\ 
 41586 & 9~12~00.287 &-64~54~06.22 &17.013 &16.899 &       &       & 12.4 & ~8.9 & \\
 45468 & 9~11~47.711 &-64~52~53.95 &17.095 &17.043 &       &       & 16.0 & ~8.4 & \\
 45560 & 9~12~25.347 &-64~52~52.13 &17.029 &16.897 &       &       & 12.7 & 11.1 & \\
 46225 & 9~13~01.673 &-64~52~40.20 &18.002 &17.968 &       &       & ~5.3 &      & \\
 47058 & 9~12~23.577 &-64~52~27.15 &17.344 &17.261 &       &       & ~7.0 &      & \\
 48260 & 9~11~43.336 &-64~52~08.19 &16.577 &16.469 &15.661 &       & 13.6 & 11.4 & \\
 48803 & 9~12~23.235 &-64~51~59.61 &16.502 &16.290 &14.324 &  114  & 18.1 & 15.4 & \\ 
 50078 & 9~12~20.455 &-64~51~40.34 &18.101 &18.014 &       &       & ~3.1 & ~6.9 & \\
 53523 & 9~11~44.322 &-64~50~45.47 &16.916 &16.783 &15.626 &       & 11.7 & ~8.9 & \\
 54675 & 9~11~46.803 &-64~50~27.35 &17.512 &17.516 &       &       & ~6.3 & 10.9 & \\
 56118 & 9~12~10.933 &-64~50~02.59 &17.103 &17.009 &       &       & 11.2 & 12.0 & \\ 
 56240 & 9~12~13.574 &-64~50~00.50 &17.092 &16.988 &       &       & 11.1 & 12.1 & \\
 56998 & 9~12~10.271 &-64~49~47.54 &17.311 &17.248 &       &       & 11.3 &      & \\ 
 58322 & 9~12~34.081 &-65~01~05.81 &17.763 &17.683 &       &       & ~5.1 & ~1.0 & \\
\hline
\multicolumn{10}{c}{Red HB}\\
\hline
  7858 & 9~12~25.778 &-64~49~11.09 &17.192 &16.328 &13.988 & ~95.5 & 18.0 & 21.9 & \\
  8288 & 9~12~02.282 &-64~48~55.90 &17.206 &16.334 &14.003 & 105.8 & 19.0 & 18.7 & \\
  8342 & 9~12~21.241 &-64~48~53.89 &17.101 &16.298 &14.188 & 111.2 & 20.4 & 19.6 & \\
  8982 & 9~12~48.755 &-64~48~24.15 &17.185 &16.324 &13.986 & 107.3 & 22.9 & 18.9 & \\
  9792 & 9~12~38.353 &-64~47~38.59 &17.217 &16.348 &13.900 & 105.3 & 23.4 & 17.0 & \\
 10026 & 9~12~18.448 &-64~47~22.11 &17.138 &16.276 &14.011 & 107.2 & 23.7 &      & \\
 10344 & 9~11~44.067 &-64~46~52.29 &17.148 &16.248 &14.003 & ~30.7 & 23.6 & 18.7 & Not member \\
 10377 & 9~11~48.520 &-64~46~48.72 &17.151 &16.314 &14.086 & 102.3 & 28.0 & 19.9 & \\
 10421 & 9~12~06.487 &-64~46~44.20 &17.126 &16.247 &13.902 & ~94.2 & 23.9 & 18.4 & \\
 10484 & 9~12~00.239 &-64~46~36.50 &17.094 &16.225 &13.958 & 106.9 & 26.8 & 21.5 & \\
 10574 & 9~11~50.176 &-64~46~26.57 &17.082 &16.258 &14.237 & ~55.3 & 24.7 & 21.4 & Not member \\
 10769 & 9~12~15.249 &-64~46~03.65 &17.112 &16.279 &14.109 & ~99.2 & 30.6 &      & \\
 10878 & 9~12~17.186 &-64~45~47.52 &17.193 &16.334 &13.967 & ~98.2 & 25.4 & 17.2 & \\
 11761 & 9~11~46.147 &-64~41~58.73 &16.980 &16.172 &13.962 & 100.7 & 20.3 & 24.5 & \\
 13128 & 9~11~10.061 &-64~49~37.54 &17.199 &16.345 &14.022 & 102.0 & 24.9 & 25.8 & \\
 13551 & 9~11~40.153 &-64~49~04.69 &17.164 &16.356 &14.049 & 108.8 & 13.6 & 20.1 & \\
 14280 & 9~11~06.203 &-64~47~53.46 &17.117 &16.229 &14.157 & ~11.4 & 11.1 & 20.5 & Not member \\
 15084 & 9~11~10.387 &-64~45~51.64 &17.073 &16.257 &14.024 & 100.1 & 12.3 & 19.4 & \\
 15413 & 9~11~40.070 &-64~44~41.60 &17.133 &16.308 &14.349 & ~41.6 & 20.5 & 21.0 & Not member \\
 15837 & 9~11~26.451 &-64~42~41.06 &17.081 &16.261 &14.286 & ~98.4 & 22.3 & 19.8 & \\
 16122 & 9~11~27.535 &-64~40~48.86 &17.197 &16.302 &14.283 & ~51.5 & 13.2 &      & Not member \\
 29702 & 9~11~18.545 &-64~57~19.91 &17.110 &16.292 &14.404 & ~39.4 & 13.6 & 13.2 & Not member \\
 29874 & 9~11~38.076 &-64~56~48.77 &17.159 &16.261 &13.968 & ~96.1 & 17.5 & 15.8 & \\
 30014 & 9~10 26.102 &-64~56~25.55 &17.031 &16.212 &14.143 & ~96.1 & 13.1 & 12.0 & \\
\hline
\end{tabular}
\end{scriptsize}
\end{table*} 

\setcounter{table}{0}
 
\begin{table*}[htb]
\centering
\caption[]{Basic data for program stars (Cont.)}
\begin{scriptsize}
\begin{tabular}{cccccccccc}
\hline
  Star & RA (J2000) & Dec (J2000) & $B$ & $V$ & $K$ & $V_r$& S/N& S/N& Note \\
       &            &             &(mag)&(mag)&(mag)&(km/s)&HR12&HR19&      \\
\hline
 30476 & 9~10 32.738 &-64~55~27.53 &17.205 &16.341 &14.326 & ~20.9 & 14.4 & ~6.7 & Not member \\
 31495 & 9~11~10.700 &-64~53~59.04 &17.069 &16.265 &14.147 & 101.2 & 23.5 & 18.3 & \\
 32580 & 9~11~10.784 &-64~52~50.07 &17.087 &16.239 &13.913 & ~97.0 & 23.5 & 15.6 & \\
 33521 & 9~10 33.897 &-64~51~56.22 &17.100 &16.263 &14.207 & ~47.6 & 17.5 & 12.5 & Not member \\
 34706 & 9~11~11.355 &-64~50~49.22 &17.148 &16.306 &14.081 & ~96.0 & 23.4 & 20.1 & \\
 34911 & 9~11~02.365 &-64~50~35.78 &17.140 &16.266 &13.930 & ~92.5 & 21.7 & 14.6 & \\
 34953 & 9~11~39.265 &-64~50~33.62 &17.222 &16.341 &13.340 & 117.3 & 22.0 & 17.2 & \\
 35212 & 9~11~13.105 &-64~50~16.45 &17.120 &16.315 &14.181 & 100.1 & 17.5 & 28.3 & \\
 35290 & 9~11~05.628 &-64~50~09.90 &17.241 &16.358 &14.052 & ~97.6 & 14.7 &      & \\
 36347 & 9~11~03.699 &-65~01~47.67 &17.069 &16.234 &14.407 & ~28.9 & 17.2 & 13.3 & Not member \\
 36638 & 9~12~27.005 &-64~59~42.31 &16.996 &16.189 &14.091 & ~97.3 & 21.8 & 17.0 & \\
 36731 & 9~11~48.708 &-64~59~21.02 &17.060 &16.204 &14.045 & 104.0 & 23.8 & 15.6 & \\
 36742 & 9~11~55.924 &-64~59~18.54 &17.101 &16.290 &14.462 & 118.5 & 21.5 & 13.0 & Not member \\
 37086 & 9~11~47.948 &-64~58~04.76 &17.055 &16.257 &14.000 & ~-5.0 & 19.8 & 18.4 & Not member \\
 37747 & 9~11~48.167 &-64~56~45.31 &17.104 &16.289 &13.990 & ~94.7 & 19.5 & 11.5 & \\
 39743 & 9~11~58.834 &-64~54~57.92 &17.162 &16.302 &13.994 & 105.5 & 22.1 & 17.8 & \\
 41995 & 9~12~47.108 &-64~53~56.37 &17.126 &16.333 &13.566 & ~96.3 & 21.8 & 17.4 & \\
 49327 & 9~12~52.349 &-64~51~51.06 &17.085 &16.245 &13.977 & 104.6 & 23.7 & 16.9 & \\
 52411 & 9~12~51.438 &-64~51~02.46 &17.060 &16.187 &13.795 & ~97.5 & 20.1 & 18.7 & \\
 56969 & 9~12~30.387 &-64~49~47.78 &17.175 &16.340 &14.063 & ~92.9 & 21.5 & 19.8 & \\
 57999 & 9~11~49.570 &-65~03~29.51 &17.162 &16.280 &14.097 & ~16.0 & 15.9 & ~4.0 & Not member \\
 58106 & 9~11~43.156 &-65~02~40.83 &17.030 &16.209 &14.082 & 108.8 & 18.7 & 14.1 & \\
 59879 & 9~13~21.297 &-64~56~53.00 &17.089 &16.229 &13.928 & ~96.5 & 16.2 & 17.4 & \\
 60283 & 9~13~21.880 &-64~56~03.05 &17.128 &16.221 &13.966 & 101.1 & 18.4 & 21.7 & \\
 62438 & 9~13~56.774 &-64~51~39.58 &17.090 &16.217 &14.192 & ~31.5 & 10.7 & 19.3 & Not member \\
\hline
\multicolumn{10}{c}{Lower RGB}\\
\hline
   111 & 9~13~43.797 &-64~49~11.96 &17.959 &17.006 &14.552 & ~80.9 & 10.5 & 16.3 & \\
   273 & 9~13~40.612 &-64~48~47.20 &18.083 &17.077 &14.757 & ~~3.1 & ~5.6 & 12.9 & Not member \\
  9505 & 9~12~48.373 &-64~47~56.23 &18.037 &17.020 &14.318 & 107.1 & 11.5 & 16.2 & \\
 10162 & 9~12~45.713 &-64~47~09.20 &17.782 &16.724 &14.049 & 105.7 & 18.9 &      & \\
 10273 & 9~12~25.930 &-64~46~59.52 &18.221 &17.227 &14.465 & 113.5 & 14.3 & 12.1 & \\
 11925 & 9~12~22.769 &-64~41~06.29 &17.814 &16.800 &14.209 & 198.3 & 15.9 &      & Not member \\
 14174 & 9~10 48.268 &-64~48~06.23 &17.967 &16.978 &14.435 & 104.8 & 12.0 &      & \\
 14713 & 9~11~05.928 &-64~46~56.55 &17.727 &16.742 &14.067 & 107.0 & 15.6 &      & \\
 23937 & 9~10 17.427 &-64~52~27.52 &18.206 &17.259 &14.712 & ~10.3 & ~9.3 & ~8.5 & Not member \\
 29619 & 9~11~40.275 &-64~57~42.00 &17.821 &16.799 &14.420 & ~14.1 & 10.7 & 12.1 & Not member \\
 30526 & 9~10 53.972 &-64~55~22.66 &18.020 &17.068 &14.747 & ~14.3 & ~7.2 & ~8.0 & Not member \\
 30550 & 9~11~25.095 &-64~55~20.18 &17.723 &16.690 &14.058 & 102.9 & 10.3 &      & \\
 30740 & 9~11~16.898 &-64~54~59.59 &18.182 &17.240 &14.878 & 101.2 & ~7.3 & 10.9 & \\
 33684 & 9~10 59.201 &-64~51~48.84 &18.094 &17.095 &14.588 & 110.8 & 13.4 &      & \\
 34075 & 9~11~32.786 &-64~51~27.40 &17.991 &16.953 &14.274 & ~52.6 & 14.2 &      & Not member \\
 36283 & 9~11~17.389 &-65~02~15.93 &18.267 &17.245 &15.000 & ~57.2 & ~9.6 & ~9.6 & Not member \\
 36517 & 9~11~07.907 &-65~00~31.64 &17.581 &16.608 &14.277 & ~-3.4 & 14.2 & ~7.8 & Not member \\
 36833 & 9~12~21.447 &-64~58~56.24 &17.925 &16.978 &14.860 & ~53.0 & ~6.7 & 10.6 & Not member \\
 36928 & 9~12~38.652 &-64~58~36.76 &18.281 &17.275 &14.771 & -25.9 & 10.9 &      & Not member \\
 37188 & 9~13~03.884 &-64~57~48.84 &18.243 &17.247 &14.795 & ~46.0 & 10.2 & ~6.7 & Not member \\
 37743 & 9~12~15.973 &-64~56~45.45 &18.087 &17.102 &14.509 & ~95.3 & 10.6 &      & \\
 38530 & 9~12~11.733 &-64~55~50.51 &17.625 &16.602 &13.866 & ~88.2 & 15.6 &      & \\
 39189 & 9~13~01.121 &-64~55~17.94 &17.623 &16.662 &14.051 & 100.9 & 17.4 & 16.0 & \\
 40873 & 9~12~44.562 &-64~54~23.35 &17.675 &16.669 &13.961 & ~91.1 & 17.2 &      & \\
 42578 & 9~12~37.677 &-64~53~45.07 &18.261 &17.287 &14.999 & ~-6.3 & 11.0 & 10.5 & Not member \\
 44947 & 9~12~17.666 &-64~53~02.46 &18.200 &17.264 &       & ~92.5 & 12.7 &      & \\
 48667 & 9~12~51.544 &-64~52~01.25 &17.761 &16.730 &14.145 & ~50.0 & 16.4 & 16.8 & Not member \\
 50027 & 9~12~24.989 &-64~51~41.03 &18.223 &17.244 &14.799 & 108.9 & 12.1 & 11.6 & \\
 51996 & 9~12~22.779 &-64~51~09.59 &17.972 &16.978 &14.450 & 102.1 & 12.9 & 14.7 & \\
 55130 & 9~12~38.880 &-64~50~19.34 &17.668 &16.648 &14.008 & ~88.3 & 13.8 &      & \\
 57250 & 9~11~56.043 &-64~49~43.17 &17.590 &16.580 &13.829 & 110.8 & 18.6 &      & \\
 58411 & 9~12~09.748 &-65~00~29.04 &18.045 &17.077 &14.760 & ~90.0 & ~8.4 & ~9.0 & \\
 58515 & 9~12~15.966 &-65~00~00.61 &17.968 &16.975 &14.482 & ~89.7 & 10.9 &      & \\
 60828 & 9~13~34.205 &-64~54~56.58 &17.646 &16.613 &14.211 & ~25.1 & ~9.7 & 12.0 & Not member \\
 63132 & 9~13~09.746 &-64~50~15.15 &17.719 &16.670 &14.019 & 105.8 & 14.6 & 25.1 & \\
\hline
\end{tabular}
\end{scriptsize}
\label{t:tab1}
\end{table*}

\begin{center}
\begin{figure}
\includegraphics[width=8.8cm]{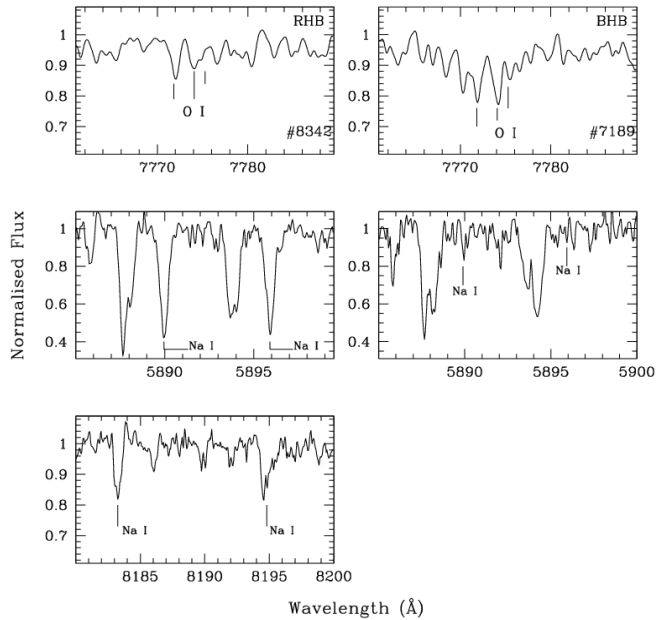}
\caption{Portions of the spectra of the RHB star \#8342 (left column)
and of the BHB star \#7189 (right column). From top to bottom: the
regions including the O~I triplet at 7771-74~\AA, the Na D lines at
5890-96~\AA, and the Na doublet at 8183-94~\AA. The spectra in the
region of the O~I triplet have been smoothed by a Gaussian with FWHM=0.3~\AA. 
The Na D interstellar lines are clearly visible in the median panels. }
\label{f:fig1}
\end{figure}
\end{center}

\begin{center}
\begin{figure}
\includegraphics[width=8.8cm]{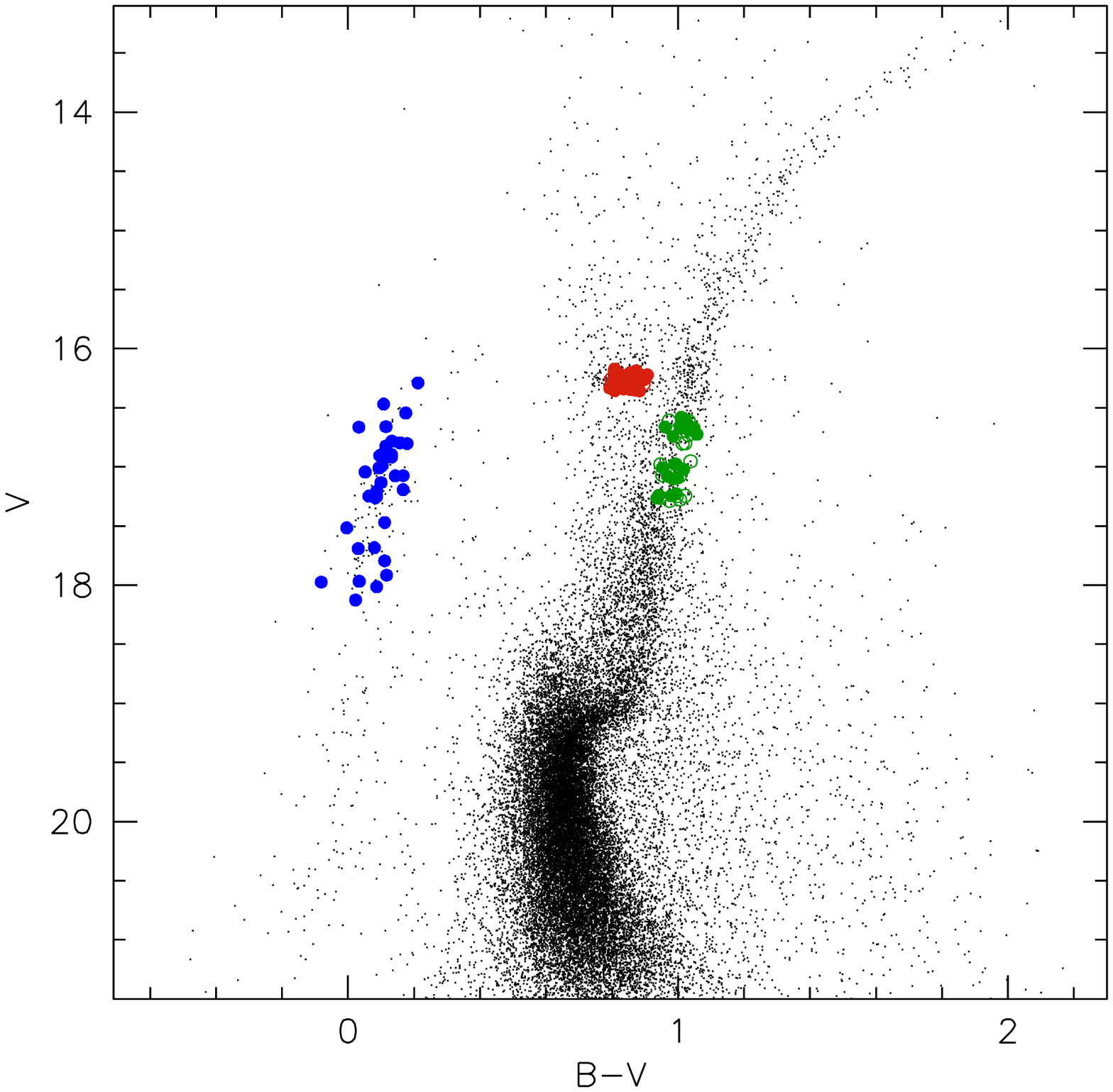}
\caption{Colour   magnitude    diagram   of    the   inner
$1\farcm5\le~R~\le10.5$  region   of  NGC2808.   Filled   circles  are
confirmed cluster  members, while open circles  (13 and 14  red HB and
red giants respectively) are non cluster members. Different colours are used for
stars on the RGB (green), RHB (red), and BHB (blue).
Stars not observed are shown as dots. }
\label{f:fig2}
\end{figure}
\end{center}

\section{Observation}

We used the high multiplex capability of GIRAFFE at VLT (Pasquini et al. 2004)
to acquire spectra for 49 stars along the red HB (RHB) and 36 on the blue HB 
(BHB) of NGC~2808. This was the maximum number of fibres that we were able to allocate 
to stars in these evolutionary phases given the practical limitations caused by the 
finite size of the fibre heads\footnote{All stars were chosen to be free from any 
companion closer than 2 arcsec and brighter than $V+2$~mag, where $V$ is the 
target magnitude.}. In addition, 34 fibres were used to point faint stars along 
the RGB (below the RGB bump), the remaining ones being used to acquire sky spectra. The median spectra 
from these last fibres were subtracted from those used for the stars. This was of 
particular relevance here, because the
observed stars are typically very faint. Two spectral configurations were
used, HR12 and HR19, providing high-resolution spectra including the strongest
features of O~I (the IR triplet at 7771-74~\AA) and Na~I (the resonance D doublet
at 5890-96~\AA, as well as the subordinate strong doublet at 8183-94~\AA)
accessible from ground. The O~I triplet and the D~lines of Na~I are the only
lines of these two elements that can be used to determine O and Na abundances
without a prohibitively observing time for NGC~2808. A few lines of Mg, Al, Si,
Ca, Fe, and Ba were also included in the selected observing ranges.

Our programme was executed in service mode, and not all planned observations
were actually carried out, so that in total we had 1.5\,hrs ($2\times 45$\,min exposures) and 
6.15\,hrs ($7\times 53$\,min) of observation with the gratings HR12 and HR19, 
respectively. The S/N of the summed spectra is typically $\sim 20$ for the RHB 
stars. For the BHB stars, spectra with grating HR12 have S/N$\sim 15$\ and those
with grating HR19 S/N$\sim 10$. The spectra were reduced using the 
ESO FLAMES GIRAFFE pipeline version 2.8.7. Sky subtraction, combination of 
individual exposures for each star, translation to rest-frame and continuum 
tracing were performed within IRAF
\footnote{IRAF is distributed by the National Optical Astronomical
Observatory, which are operated by the Association of Universities
for Research in Astronomy, under contract with the National Science
Foundation}. Telluric
lines were removed from the longest wavelength spectra by subtracting the
average spectrum of those BHB stars for which the observations have $S/N>8$.
This combined spectrum has a $S/N\sim 50$ (much higher than the $S/N$ of the
individual programme stars), and was obviously taken with the same airmass than
the programme star, so that the excision of the telluric lines turned out to be
excellent. Examples of spectra are shown in Figure~\ref{f:fig1}.

Figure~\ref{f:fig2} shows the location of the programme stars on the colour magnitude
diagram of NGC~2808. Our NGC2808 ground-based photometric catalogue (see  Momany  et
al. 2004) consists of $UBV$ observations obtained at the Wide-Field
Imager (WFI) mounted on the 2.2m  ESO-MPI (La Silla, Chile). The WFI
catalogue has a total field of view of
$34^{\prime}\times~33^{\prime}$.  Photometric data for the programme stars are listed
in Table~\ref{t:tab1}. The $K$\ magnitudes are from the 2MASS point source catalogue
(Skrutskie et al. 2006). Note that no information on membership of the 
programme stars to the cluster was available prior to the observations. The 
stars selected for observations lie close to the mean loci of the colour 
magnitude diagram of NGC~2808. Virtually no field contamination is expected 
for the BHB, while some field interlopers may be present in our RHB sample. 
We then determined their membership from the radial velocities (also listed
in Table~\ref{t:tab1}), because the fairly high radial velocity of NGC~2808 
($+101.6\pm 0.7$~km\ts s$^{-1}$, Harris 1996) ensured that very few field stars with similar 
radial velocities were to be expected. Thirty-seven RHB stars turned out to be cluster 
members based on this criterion. A close inspection
revealed that most of the cluster members are grouped in a very narrow region
of the colour magnitude diagram, while the field stars are typically slightly
bluer and fainter than the real RHB stars. After the abundance analysis, we
excluded another star (\#36742) because its Fe abundance is much larger than
typical for stars in NGC~2808. Note that the radial velocity and the K magnitude of this star
only marginally agree with those of the NGC~2808 RHB members.
A similar analysis for the RGB stars indicated that 22 stars are cluster members.
The average radial velocity of the 36 RHB stars is $+101.0\pm 0.9$~km\ts s$^{-1}$,
in agreement with the value listed by Harris (1996). The r.m.s. scatter is
5.7~km\ts s$^{-1}$. The average radial velocity of the lower RGB (21 stars)
is $+99.9\pm 3.0$~km\ts s$^{-1}$, with an r.m.s. scatter of 
9.2~km\ts s$^{-1}$. The difference between the r.m.s. scatter
obtained for RHB and RGB stars is not significant.

Not unexpectedly, most of the BHB stars of NGC~2808 turned out to be hotter
than 11,500~K, with only six stars cooler than this limit. In the following, we
will present our result for these last stars and for the RHB member stars. 
Results for the other stars will be presented elsewhere.

We also measured radial velocities for the BHB stars cooler than the
Grundahl $u-$jump, though with fairly large errors because they are based on
the O~I triplet lines alone. Three of these stars are in common with the study of Recio-Blanco 
et al. (2004). These are 48003=2333, 10449=3949, and 37289=3841 (the first number
is our identification, the second that from Recio-Blanco et al. 2004);
however, we have radial velocity only for two of them. The comparison is fairly
good, with a difference of $2\pm 3$~km\ts s$^{-1}$. 

Various authors found that horizontal branch stars rotate with velocities up to 
a few tens of km\ts s$^{-1}$ for BHB stars (Peterson et al. 1995; Behr et al. 2000a, 2000b) 
and slightly less for RHB ones (Carney et al. 2008). No really fast rotator was found
among the 16 BHB stars observed by Recio-Blanco et al. (2004) in NGC2808, where the
maximum rotation velocity was $13\pm 4$~km\ts s$^{-1}$\ for their star 3841, which is our star
37289. We have checked our sample for fast rotators. To this aim, we considered the FWHM 
of the fit to the cross-correlation peak obtained using as a template a spectrum 
of star of similar atmospheric parameter taken with the same instrument. To derive 
v$\sin{i}$ from the FWHM the contribution caused by the instrumental profile, macro- 
and micro-turbulence, and thermal Doppler broadening should be taken into account, 
both for the stars and the template, and the relation should be calibrated (see 
Lucatello \& Gratton 2003). However, the FWHM of the cross correlation of spectra 
of stars with similar atmospheric parameters with the same template gives a 
qualitative indication of the rotational rate. To this aim we used the spectra taken 
with grating HR12, because in HR19 the number of stellar lines is much smaller and the 
spectra are dominated by telluric lines.

Among red HB stars we found no evidence of a considerable variation in the measured 
FWHMs, which range from $\sim$28 to $\sim$40~km\ts s$^{-1}$. While these values cannot
exclude the presence of stars rotating as fast as v$\sin i \sim$5-7~km\ts s$^{-1}$ (typical
values for field RHB stars observed by Carney et al. 2008), just the 
accounting of the instrumental profile of template and object stars at the spectral 
resolution used (R$\simeq$18700) leaves no room for high rotational velocities. We 
cannot draw any conclusions concerning blue HB stars: the paucity of lines 
combined with the quite low signal-to-noise ratio makes the estimates of FWHM faily
uncertain as well as highly affected by noise, residuals of telluric subtraction, etc.

\section{Analysis}

\subsection{Atmospheric parameters}

The programme stars cover a wide range in effective temperature. It is therefore difficult 
to mantain a uniform criterium to derive atmospheric parameters. In addition,
the low S/N and the limited spectral coverage of the spectra prevent a full
spectroscopic analysis. The adopted procedure was the following. 

For the RHB and RGB stars, effective temperatures 
were derived from the $B-V$\ and $V-K$\ colours, using the calibration of Alonso 
et al. (1999, with the erratum of Alonso et al. 2001). The colours were dereddened 
using the $E(B-V)$\ values from the updated on-line version of the Harris (1996) 
catalogue and the $E(V-K)/E(B-V)$\ value from Cardelli et al. (1989). The calibrations 
require input values for the metallicity [A/H]. We adopted the value obtained by
Carretta et al. (2006). We assign weight 4 to the $B-V$\ colours, and 1 to the $V-K$\ ones, because the
programme stars are very faint for the 2MASS observations. For the blue HB stars 
we used the $(B-V) - T_{\rm eff}$\ calibration by Kurucz\footnote{See kurucz.harvard.edu.}.

Errors are mainly caused by uncertainties in the colours. For the RHB stars, an error
of 0.01 mag in $B-V$\ causes an error of 25~K in the effective temperature; a slightly smaller error
of 20~K is obtained for red giants. A similar photometric error for the BHB stars causes 
an error of $\sim 300$~K. We assumed errors of 50 and 300~K as representative 
values for the internal errors in the temperatures. Systematic errors caused by scale errors 
or incorrect parameters for the cluster are likely larger. However, they are not really 
important in our discussion.

The surface gravities were obtained from the masses, luminosities, and effective temperatures. For 
the masses we adopted values of 0.9, 0.7, and 0.5~$M_\odot$\ for stars on the RGB,
RHB, and BHB (see Gratton et al. 2010, for a discussion of
adequate values for the different sequences). Note that these values are not very
critical, because a variation of 0.1~$M_\odot$\ changes gravities by 0.07 and 0.10 dex
for RHB and BHB stars. The effects on [Na/Fe] are of 0.11 and 0.08 dex, and those on
[O/Fe] 0.02 and 0.13 dex, for BHB and RHB respectively.
The bolometric corrections needed in these derivations
were obtained using calibrations consistent with those used for the effective temperatures (Alonso et
al. 1999 for the red giant and RHB stars, and Kurucz for the BHB
stars). The distance moduli used were taken from the Harris catalogue.

Errors in gravities are very small. The assumption about masses is likely correct within 10\% (0.04~dex
error in the gravities), while that on the effective temperature and luminosity causes errors in
gravities not larger than $\sim 2$\% for the RHB stars and red giants, and $\sim 12$\%
for the BHB stars. The error in gravities is then not larger than 0.05~dex for the
cool stars, while it may be 0.10~dex for the warm ones.

The same metal abundance of [A/H]=-1.14 and microturbulence velocity of 1.8~km\ts s$^{-1}$
were adopted for all RHB stars. The metal abundance is the average value determined
in Carretta et al. (2006), and agrees well with the average [Fe/H]=$-1.18\pm 0.01$
(r.m.s of 0.07~dex) we determined from the 36 RHB stars. The adopted microturbulence 
velocity sets to zero the average of the trends of Fe abundances with expected line 
strength (see Magain 1984 for a justification of this approach). We prefer to adopt 
this average value rather than values appropriate for each star because we were able to measure 
equivalent widths (EW) only for a limited number of Fe lines (in the range 6-16) for each 
star, with fairly large errors ($\sim 13$~m\AA, evaluated from the r.m.s. of measures
for the same line across the RHB sample), and over a limited range of EWs. The
uncertainties in the individual slopes, which agree with the observed r.m.s. scatter, 
would translate into a large error bar of $\pm 0.5$~km\ts s$^{-1}$. This is likely
more than the expected star-to-star scatter, which we estimated to be 
$\sim 0.3$~km\ts s$^{-1}$\ in analogy with the RGB stars studied by Carretta et al. (2009a).
We will adopt this last value as the error bar of our microturbulent velocities.
We were unable to derive microturbulent velocities for BHB stars because too few
lines were measured. The value we adopted (2~km\ts s$^{-1}$) is in the middle of the
range usually found in previous analysis of BHB stars (Lambert et al. 1992; Behr et
al. 1999, 2000b; Kinman et al. 2000; Fabbian et al. 2005; Villanova et al. 2009;
Marino et al. 2011).

While a single Fe~II line was measured in our spectra (at 5991.38~\AA),
abundances derived from this line agree well with those obtained from the Fe~I lines:
on average we obtained [Fe~II/H]=$-1.13\pm 0.03$. This supports the choice of the atmospheric
parameters adopted throughout this paper.
 
\begin{table*}[htb]
\centering
\caption[]{Atmospheric parameters and abundances}
\begin{scriptsize}
\begin{tabular}{ccccccccccccccccc}
\hline
Star &$T_{\rm eff}$&$\log{g}$& &$[$Fe/H$]_I$& &$[$Fe/H$]_{II}$& &$[$O/Fe$]$& & &$[$Na/Fe$]$& &$[$Mg/Fe$]$&$[$Si/Fe$]$&$[$Ca/Fe$]$&$[$Ba/Fe$]$\\
     &(K) & (dex) & lines & $<>$ & rms &  & lines & $<>$ & rms & lines & $<>$ & rms & & & & \\
\hline
\multicolumn{17}{c}{Blue HB}\\
\hline
 7189 & 10020 & 3.30 &   &       &      &       & 3 &-0.23 &      & 2 &  0.44 & 0.07 &      &      &      &       \\
 8008 & 10270 & 3.33 &   &       &      &       & 2 &-0.29 &      & 1 &  0.45 &      &      &      &      &       \\
10449 &  9900 & 3.29 &   &       &      &       &   &      &      & 1 &  0.25 &      &      &      &      &       \\
32324 & 11120 & 3.37 &   &       &      &       & 2 &-0.66 &      & 2 &  0.61 & 0.07 &      &      &      &       \\
37289 & 10560 & 3.29 &   &       &      &       & 1 &-0.71 &      & 2 &  0.61 & 0.09 &      &      &      &       \\
48803 &  9060 & 3.17 &   &       &      &       & 3 &-0.75 &      & 2 &  0.65 & 0.01 &      &      &      &       \\
\hline
\multicolumn{17}{c}{Red HB}\\
\hline
 7858 & 5490 & 2.53 & 14 & -1.27 & 0.29 & -1.09 &2 & 0.54 & 0.03 & 4 & -0.15 & 0.17 & 0.16 & 0.04 & 0.61 &  0.49 \\
 8288 & 5470 & 2.52 & 15 & -1.25 & 0.32 & -0.89 &3 & 0.60 & 0.09 & 4 &  0.09 & 0.08 & 0.12 & 0.30 & 0.20 &  0.65 \\
 8342 & 5680 & 2.58 & 15 & -1.11 & 0.28 & -1.07 &2 & 0.43 & 0.04 & 4 &  0.28 & 0.20 & 0.36 & 0.35 & 0.45 &  0.52 \\
 8982 & 5490 & 2.53 & 16 & -1.20 & 0.30 & -1.20 &1 & 0.38 &      & 4 & -0.02 & 0.07 & 0.26 & 0.17 & 0.37 &  0.19 \\
 9792 & 5450 & 2.52 & 16 & -1.26 & 0.30 & -1.26 &2 & 0.79 & 0.21 & 4 &  0.06 & 0.07 & 0.04 & 0.47 & 0.41 &  0.42 \\
10026 & 5510 & 2.51 & ~6 & -1.16 & 0.18 & -1.32 &  &      &      & 2 &  0.00 & 0.17 &      &      & 0.45 &  0.56 \\
10377 & 5570 & 2.55 & 16 & -1.19 & 0.19 & -1.35 &3 & 0.67 & 0.26 & 4 & -0.04 & 0.27 & 0.09 & 0.60 & 0.15 &  0.00 \\
10421 & 5460 & 2.48 & 16 & -1.19 & 0.21 & -1.26 &3 & 0.15 & 0.02 & 4 & -0.17 & 0.37 & 0.19 & 0.49 & 0.42 &       \\
10484 & 5500 & 2.49 & 16 & -1.13 & 0.16 & -1.03 &2 & 0.54 & 0.02 & 4 & -0.21 & 0.21 & 0.07 & 0.61 & 0.54 &  0.18 \\
10769 & 5600 & 2.55 & ~6 & -1.05 & 0.07 & -1.08 &  &      &      & 2 &  0.00 & 0.02 &      &      & 0.49 &  0.18 \\
10878 & 5490 & 2.53 & 16 & -1.16 & 0.19 & -0.82 &2 & 0.49 & 0.20 & 4 &  0.07 & 0.34 & 0.09 & 0.53 & 0.41 & -0.03 \\
11761 & 5640 & 2.52 & 14 & -1.12 & 0.28 & -1.29 &3 & 0.44 & 0.16 & 3 &  0.23 & 0.30 & 0.28 & 0.32 &      &  0.15 \\
13128 & 5510 & 2.54 & 13 & -1.22 & 0.20 & -1.17 &2 & 0.18 & 0.01 & 4 & -0.18 & 0.20 & 0.11 & 0.14 & 0.35 & -0.22 \\
13551 & 5610 & 2.58 & 13 & -1.26 & 0.22 & -0.83 &3 & 0.51 & 0.14 & 4 &  0.02 & 0.21 & 0.21 & 0.29 & 0.48 &  0.45 \\
15084 & 5620 & 2.55 & 16 & -1.12 & 0.24 & -1.32 &3 & 0.47 & 0.01 & 4 &  0.11 & 0.20 & 0.35 & 0.47 & 0.63 &  0.42 \\
15837 & 5690 & 2.57 & 15 & -1.06 & 0.22 & -0.73 &2 & 0.33 & 0.14 & 4 &  0.08 & 0.38 & 0.17 & 0.39 & 0.44 &  0.12 \\
29874 & 5450 & 2.48 & 14 & -1.04 & 0.22 & -1.09 &3 & 0.64 & 0.10 & 4 & -0.09 & 0.19 & 0.53 & 0.58 & 0.35 &  0.43 \\
30014 & 5660 & 2.54 & 14 & -1.12 & 0.22 & -1.06 &3 & 0.40 & 0.13 & 4 &  0.33 & 0.33 & 0.16 & 0.42 & 0.92 &  0.29 \\
31495 & 5680 & 2.57 & 14 & -1.17 & 0.27 & -1.03 &3 & 0.39 & 0.29 & 4 &  0.34 & 0.37 & 0.33 & 0.04 & 0.47 &  0.49 \\
32580 & 5520 & 2.50 & 14 & -1.18 & 0.20 & -1.18 &2 & 0.38 & 0.19 & 4 & -0.06 & 0.33 & 0.27 & 0.38 & 0.38 & -0.19 \\
34706 & 5560 & 2.55 & 15 & -1.23 & 0.17 & -1.12 &3 & 0.54 & 0.09 & 4 & -0.08 & 0.27 & 0.10 & 0.32 & 0.59 &  0.31 \\
34911 & 5470 & 2.49 & 13 & -1.16 & 0.24 & -1.30 &2 & 0.35 & 0.37 & 4 & -0.20 & 0.25 & 0.01 & 0.64 & 0.31 &  0.29 \\
34953 & 5460 & 2.52 & 13 & -1.27 & 0.26 & -1.58 &1 & 0.69 &      & 4 & -0.34 & 0.29 & 0.01 & 0.10 & 0.65 &  0.41 \\
35212 & 5670 & 2.59 & 12 & -1.31 & 0.32 & -1.24 &3 & 0.14 & 0.05 & 4 &  0.28 & 0.28 &      & 0.17 & 0.33 &  0.18 \\
35290 & 5460 & 2.53 & ~6 & -1.29 & 0.17 & -1.02 &  &      &      & 2 & -0.04 & 0.07 &      &      & 0.10 & -0.11 \\
36638 & 5680 & 2.54 & 14 & -1.06 & 0.27 & -1.06 &2 & 0.40 & 0.23 & 4 &  0.14 & 0.35 & 0.47 & 0.47 & 0.49 &  0.74 \\
36731 & 5550 & 2.50 & 15 & -1.15 & 0.36 & -1.35 &3 & 0.51 & 0.08 & 3 & -0.12 & 0.13 & 0.28 & 0.37 & 0.45 &  0.59 \\
37747 & 5600 & 2.55 & 16 & -1.11 & 0.37 & -1.16 &2 & 0.32 & 0.01 & 4 &  0.20 & 0.29 & 0.13 & 0.50 & 0.25 &  0.22 \\
39743 & 5500 & 2.52 & 14 & -1.18 & 0.19 & -1.18 &3 & 0.32 & 0.36 & 4 & -0.01 & 0.13 & 0.20 & 0.58 & 0.54 &  0.01 \\
41995 & 5690 & 2.60 & 14 & -1.20 & 0.23 & -1.16 &3 & 0.37 & 0.26 & 4 & -0.02 & 0.19 & 0.38 & 0.34 & 0.07 &  0.07 \\
49327 & 5560 & 2.52 & 14 & -1.24 & 0.26 & -0.95 &3 & 0.56 & 0.23 & 4 & -0.26 & 0.15 & 0.14 & 0.45 & 0.49 & -0.13 \\
52411 & 5460 & 2.46 & 15 & -1.19 & 0.21 & -1.31 &3 & 0.45 & 0.15 & 4 &  0.22 & 0.22 & 0.37 & 0.34 & 0.32 &  0.47 \\
56969 & 5560 & 2.56 & 15 & -1.23 & 0.22 & -1.21 &3 & 0.46 & 0.06 & 4 & -0.10 & 0.19 & 0.34 & 0.46 & 0.54 &  0.46 \\
58106 & 5640 & 2.53 & 11 & -1.13 & 0.16 & -0.96 &1 & 0.13 &      & 4 &  0.29 & 0.06 & 0.22 & 0.42 & 0.41 &  0.28 \\
59879 & 5510 & 2.49 & 12 & -1.30 & 0.24 & -1.09 &3 & 0.65 & 0.14 & 4 & -0.13 & 0.20 & 0.14 &      & 0.41 & -0.09 \\
60283 & 5430 & 2.46 & 15 & -1.11 & 0.22 & -1.00 &3 & 0.38 & 0.11 & 4 & -0.05 & 0.21 & 0.53 & 0.27 & 0.41 & -0.21 \\
\hline
\end{tabular}
\end{scriptsize}
\label{t:tab2}
\end{table*}

Table~\ref{t:tab2} lists the effective temperatures $T_{\rm eff}$\ and 
surface gravities $\log{g}$\ we used in the analysis of the programme stars, 
as well as the abundances we obtained from our analysis for 
lines of Fe~I, Fe~II, O~I, Na~I, Mg~I, Si~I, Ca~I, and Ba~II. Abundances were
estimated from equivalent widths. The oscillator strengths for Fe 
lines were generally obtained from the VALD
database (Kupka et al. 2000). The same source was used for the Mg line at
8213.04~\AA, for four lines of Si (5948.55, 7918.38, 7932.35 and 7944.00~\AA), for
the Ca line at 5857.46~\AA, and for the Ba~II line at 5853.69~\AA. Note that
not all stars were observed with both gratings. As a consequence, we were not
able to measure the abundances of all elements in each star. For this reason
we have no O abundance for star \#10449, for instance.

For O~I and Na~I lines, the abundances include corrections due to departures from 
LTE, which are large for the transition and stars of interest here. The non-LTE 
corrections used are estimated from standard statistical equilibrium calculations. 
They include integration of the radiation transfer equation throughout the stellar 
atmosphere, with corrections to the populations of the individual levels and to the 
source function computed at all optical depth. There are uncertainties, related to 
the approximation present and to the poor knowledge of the relevant cross sections. 
In particular the treatment of collisions with H~I atoms is critical and poorly known. 
Because different recipes are used, various non-LTE estimates produce somewhat 
different results, although there is generally quite a good consensus on the sign 
and order of magnitude of the corrections, in particular for quite simple atoms like 
O~I and Na~I. For the RHB stars, we used the corrections listed by Gratton et al. 
(1999), whose cross sections for collisions with H~I atoms were tuned to reproduce 
observations of RR Lyrae stars. The corrections are $\sim 0.5$~dex for O~I and 
$\sim 0.2$~dex for Na, in both cases decreasing the LTE abundances. Very recently, new 
Na non-LTE corrections were published by Lind et al. (2011), making use of a 
more extended model atom and moreover updated data about cross sections. For the lines 
used and the combination of atomic parameters appropriate for the programme stars, these 
corrections are about 0.3 dex larger than those adopted here and this difference is 
quite uniform. Were we to adopt these corrections, the Na abundances would be about 
0.3~dex smaller. However, we prefer to keep here the older correction by Gratton et al. 
(1999) for uniformity with the red giants studied by Carretta et al. (2006). However, 
even these corrections were not computed for stars as hot as our BHB stars. In 
this case, we used the results of the statistical equilibrium calculations by Takeda 
(1997) for O~I (with corrections to O abundances in the range $1-1.5$~dex, and by 
Mashonkina et al. (2000) for Na~I (in this case the corrections are $\sim 0.5$~dex). 
We alert the reader that there might be some offsets between these different sets of 
corrections. These offsets are difficult to quantify accurately. If we 
compare non-LTE abundance corrections for O triplet lines by Takeda (1997) with those 
by Gratton et al. (1999) for stars with $T_{\rm eff}=7000$~K, however, where both are available, 
we find that the first ones are smaller by $\sim 0.1$~dex ($\sim 0.5~vs~\sim 
0.6$~dex). A similar comparison for the Na non-LTE corrections of Mashonkina et al. 
(2000) shows that these are smaller by $\sim 0.2$~dex than those by Gratton et al. (1999)
($\sim 0.4~vs~\sim 0.6$~dex). We conclude that uncertainties in the non-LTE corrections 
may be likely as large as a few tenths of a dex (see also comparisons in Lind et al. 2011, 
which are appropriate to cooler stars, however). Hence, while these uncertainties are not 
negligible, it is difficult to conceive that they may cause the main results of this paper.

The doublet of Al~I at 7835-36~\AA\ is actually in the observed spectral range.
We carefully examined the relevant region in the spectra of all RHB stars,
but we did not find any evidence for it. The line is also not detectable
in the spectrum we obtained by summing all 36~HB stars. From this last spectrum 
(which has a $S/N>100$\ at the relevant wavelength)
we derived an upper limit of 10~m\AA\ to the equivalent width for each
component of the doublet. This yields an upper limit of [Al/Fe]$<0.15$,
which is similar to the Al abundances obtained for the most Al-poor RGB stars
in NGC~2808 (Carretta et al. 2009b). No Al detection is possible in the low S/N spectra of BHB stars.

\subsection{Error analysis}

\begin{table*}[htb]
\centering
\caption[]{Sensitivity and error analysis}
\begin{tabular}{lcccccc}
\hline
Parameter &$T_{\rm eff}$&$\log{g}$&$[$A/H$]$&$v_t$& EW & Total \\
\hline
Variation& +100 K & +0.5 dex & +0.1 dex & +0.5 km/s &+10 m\AA &     \\
Error BHB& 300 K & ~0.1 dex & ~0.1 dex & ~0.5 km/s &~20 m\AA &        \\
Error RHB& ~50 K & ~0.1 dex & ~0.1 dex & ~0.3 km/s &~13 m\AA &        \\
\hline
\multicolumn{7}{c}{Blue HB}\\
\hline
$[$O/Fe$]$  &  0.025 & ~0.016 & ~0.008 & -0.129 & 0.085 & 0.226 \\ 
$[$Na/Fe$]$ &  0.060 & -0.165 & -0.003 & -0.014 & 0.109 & 0.284 \\ 
\hline
\multicolumn{7}{c}{Red HB}\\
\hline
$[$Fe/H$]$  & ~0.065 & -0.020 & -0.001 & -0.098 & 0.050 & 0.070 \\ 
$[$O/Fe$]$  & -0.095 & -0.181 & ~0.000 & -0.081 & 0.080 & 0.129 \\ 
$[$Na/Fe$]$ & ~0.095 & -0.125 & ~0.003 & -0.126 & 0.049 & 0.112 \\ 
$[$Mg/Fe$]$ & ~0.025 & -0.015 & ~0.000 & -0.022 & 0.160 & 0.209 \\ 
$[$Si/Fe$]$ & ~0.030 & ~0.005 & ~0.001 & -0.027 & 0.094 & 0.124 \\ 
$[$Ca/Fe$]$ & ~0.075 & -0.100 & ~0.000 & -0.200 & 0.158 & 0.242 \\ 
$[$Ba/Fe$]$ & ~0.056 & ~0.181 & ~0.006 & -0.169 & 0.181 & 0.260 \\ 
\hline
\end{tabular}
\label{t:tab3}
\end{table*}

Error analysis was made in the usual way, by repeating the abundance
derivation by modifying a single parameter each time. Relevant data
are given in Table~\ref{t:tab3}. The last column gives an estimate of
the total internal errors, estimated using the sensitivities listed above,
as well as the errors in the individual parameters given in lines 2 and
3 for blue and red HB stars, respectively. In general, the main source
of uncertainty is the measure of the equivalent widths, which is
not surprising considering the fairly low S/N of our spectra. In a few
cases (Fe, O, and Na) a significant contribution comes from errors in the
effective temperatures and in the microturbulent velocities.

For most elements (Fe, Mg, Si, Ca, and Ba) the dispersion of individual
values (for RHB stars alone) agrees fairly well with these estimates
of the internal errors. The case of Na and O will be discussed in the
next section.

\begin{center}
\begin{figure}
\includegraphics[width=8.8cm]{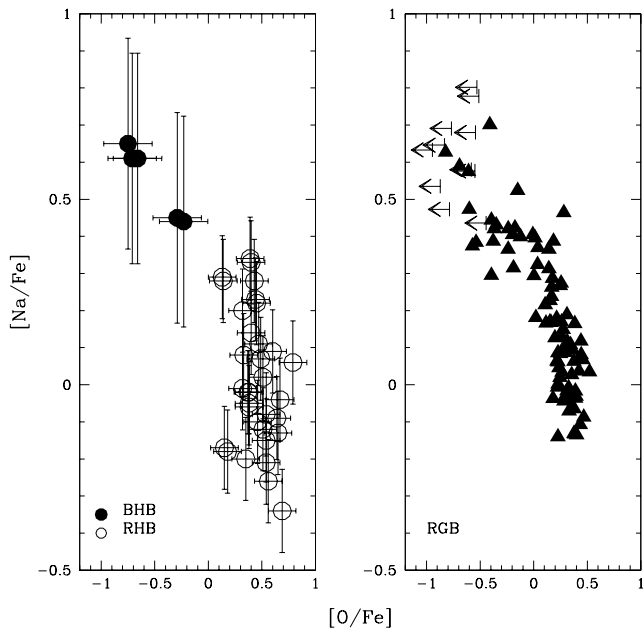}
\caption{Left panel: Na-O anticorrelation for HB stars of NGC~2808. Filled circles are
BHB stars; open circles are RHB stars. Right panel: the same, but for red giants
(from Carretta et al. 2006). }
\label{f:fig3}
\end{figure}
\end{center}

\begin{center}
\begin{figure}
\includegraphics[width=8.8cm]{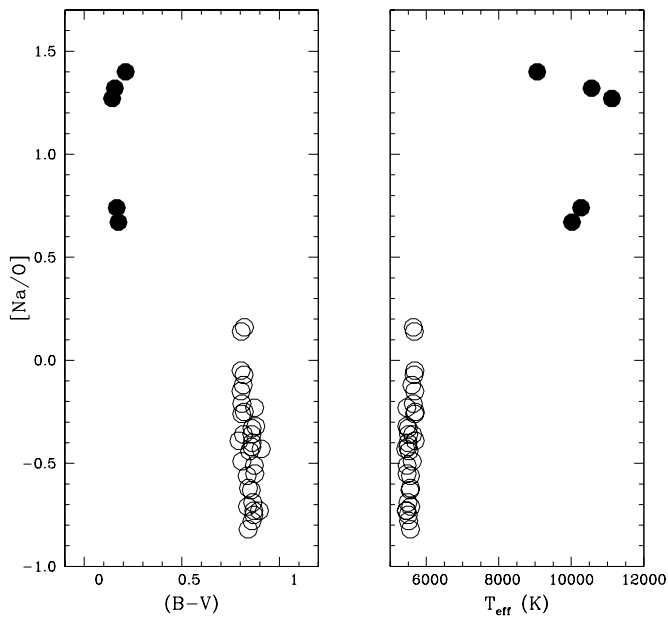}
\caption{Left: Run of the [Na/O] abundance ratio with $B-V$\ colour along the
HB of NGC~2808. Filled circles and open circles are for blue and red HB
stars, respectively. Right: the same, but with $T_{\rm eff}$ rather than
colour.}
\label{f:fig4}
\end{figure}
\end{center}

\section{The Na-O anticorrelation along the HB of NGC~2808}

Figure~\ref{f:fig3} shows the Na-O anticorrelation we obtained for the HB stars of
NGC~2808. Different symbols are used for blue and red HB stars. For
comparison we also plotted the Na-O anticorrelation for red giants
by Carretta et al. (2006). We recall that the extreme blue HB stars, 
hotter than 11,500~K, are not considered here, because their surface abundance 
is not related in a simple way to their original composition. The moderately 
BHB stars for which we may provide sensible Na and O abundances are all
O-poor and Na-rich, with average values of [O/Fe]=$-0.59\pm 0.12$\ and
[Na/Fe]=$+0.44\pm 0.06$. The r.m.s. scatter of individual values of 
0.24~dex and 0.15~dex agree with the internal errors of Table~\ref{t:tab3},
so that we cannot conclude about any internal scatter in this small sample of six 
stars, even if there is a hint that the stars might be distributed into
two groups. We also notice that systematic errors caused by, e.g. the corrections
for non-LTE effects (which are rather large for the O lines) are certainly
much larger than the error bars given above for the average abundances,
which stems from the dispersion of individual values alone. However, the
comparison of the Na and O abundances for these stars with those obtained
for red giants suggests that they descend from the moderately O-poor red
giants.

On the other hand, all RHB stars are much richer in O than the BHB
stars, with positive values of [O/Fe]. These stars span a fairly large
range of Na abundances, and the r.m.s. scatter of [Na/Fe] values of 
$0.18\pm 0.03$~dex appears to be significantly larger than the observational
error of 0.11~dex. While not extremely significant, there is a clear hint
for a Na-O anticorrelation. The correlation coefficient is $r=-0.34$\ over 31
stars, which has less than 5\% chance of being a random effect. In addition,
we notice that there is a very significant correlation between the Na
abundances obtained from the resonance D-doublet and the subordinate one at 
8183-94~\AA. This suggests that the red HB stars have not a single value
of Na and O abundances, with only observational scatter; instead they seem
to include at least two different populations. This means that only a fraction of
the RHB stars of NGC~2808 descends from the primordial population (see
Carretta et al. 2010, for a definition of primordial population). The
comparison with red giants shows that there is a very good correspondence
between the RHB stars and the O-rich sequence in NGC~2808; we incidentally
note here that also O-rich red giants display a distinct spread in Na
abundances, so that many of them are classified as belonging to the
intermediate population according to the definition by Carretta et al. (2009a).

While we cannot determine directly the He abundances
for the various populations on the HB of NGC~2808, we may propose values
that are compatible with their colours and magnitudes. This matter has been
considered in depth in previous investigations. For instance, D'Antona et al. 
(2005) considered three populations with Y=0.24, Y=0.26-0.29, and Y=0.40,
with respectively 50, 30, and 20\% of the stars, and found that this mix
is able to reproduce both the main sequence and the HB of NGC~2808. 
The first group corresponds to the RHB, the second to the BHB, and the
third to extreme BHB (not sampled in this paper). The same proportions
are found for P, I, and E stars along the RGB by Carretta et al. (2010).
This suggests the identification of the He-poor stars and RHB with the
P-stars, of the He-intermediate and BHB stars with the I-stars, and
of the He-rich and extreme BHB ones with the E-stars. As expected from 
this scenario, we indeed found a clear correlation between the Na/O 
abundance ratio and colour and temperature along the HB in NGC~2808 
(see Figure~\ref{f:fig4}). This is obvious when comparing the O-poor, 
Na-rich BHB with the O-rich RHB stars. While we were unable to analyse 
the warmer stars along the HB of NGC~2808, we think this is a strong 
circumstantial confirmation of the scenario where the location 
of stars along the HB is determined also by their helium content. 
However, because there is a hint for a trend of Na and O abundances
with colour also among the red HB stars alone, we suggest that there 
are more than three stellar populations in NGC2808, and that the He-poor
group itself divides into further groups. In order to produce the
distribution in colours of the stars, the helium abundance differences
among these groups should be small ($\leq 0.01$). However, the presence
of significant variations in Na content has implications in a scenario
for the formation of this cluster.

\section{Conclusions}

We have obtained
spectra for several tens stars on the HB of NGC~2808, which is the prototype 
mono-metallic, multiple stellar population cluster. Several red HB stars
were observed, and six on the blue HB cooler than the gap at 11,500 K
studied by Grundahl, Caloi, and Behr. We found that as expected, the blue HB stars are
O-poor and Na-rich, and the red HB ones are O-rich and Na-poor. This strongly
supports the identification of the BHB as the progeny of the intermediate MS,
and of the RHB with the red MS, made by D'Antona et al. (2005; no reliable
abundances could be obtained for the extremely blue HB stars, which should
be the progeny of the blue MS stars). As additional
information, we found that some Na-O anticorrelation can be found also among
the RHB stars alone, suggesting that this sequence which includes almost
half of the HB stars in NGC2808 itself, is made of at least two distinct groups
of stars. Hence there are more than three stellar populations in NGC2808.

\begin{acknowledgements}
This publication makes use of data products from
the Two Micron All Sky Survey, which is a joint project of the
University of Massachusetts and the Infrared Processing and Analysis
Center/California Institute of Technology, funded by the National
Aeronautics and Space Administration and the National Science
Foundation. This research has made use of the NASA's Astrophysical
Data System. This research has been funded by PRIN INAF
"Formation and Early Evolution of Massive Star Clusters".
We thank an anonymous referee for his/her useful suggestions.

\end{acknowledgements}

\end{document}